\documentclass[12pt]{article}
\usepackage{graphicx}
\usepackage{amsmath}
\usepackage{amsbsy}
\usepackage{amsfonts}
\usepackage{amsthm}
\begin{document}

\title{Entanglement: A myth introducing non-locality in any quantum theory}

\author{Athanasios Prikas}

\date{}

\maketitle

Physics Department, National Technical University, Zografou Campus,
157 80 Athens, Greece.\footnote{e-mail: aprikas@central.ntua.gr}

\begin{abstract}
The purposes of the present article are: a) To show that
non-locality leads to the transfer of certain amounts of energy and
angular momentum at very long distances, in an absolutely strange
and unnatural manner, in any model reproducing the quantum
mechanical results. b) To prove that non-locality is the result only
of the zero spin state assumption for distant particles, which
explains its presence in any quantum mechanical model. c) To
reintroduce locality, simply by denying the existence of the zero
spin state in nature (the so-called highly correlated, or EPR
singlet state) for particles non-interacting with any known field.
d) To propose a realizable experiment to clarify if two remote (and
thus non-interacting with a known field) particles, supposed to be
correlated as in Bell-type experiments, are actually in zero spin
state.
\end{abstract}

PACS number(s): 03.65.-w, 03.65.Ta, 03.65.Ud

\newpage

\section{Introduction, or what is non-locality}

Non-locality, \cite{EPR,Bell}, has been investigated by several
authors. For some recent reviews, see
\cite{entanglement1,entanglement2} for quantum entanglement,
\cite{hidden01,hidden1,hidden2,hidden3} for the non-locality problem
from the hidden variables quantum theory point of view, or
\cite{information} for some applications in quantum information
theory. Our purpose, in the present part of the article, is to prove
that two or more correlated particles, even when they are unable to
interact with a certain Hamiltonian (i.e.: when they are at great
distance, even with walls of Pb between them, even when every
particle with its measuring devices is entrapped in rooms deep
beneath the surface of the earth), they exchange energy and angular
momentum, and this is what I call "non-locality".

We will also prove that every quantum theory, orthodox or of hidden
variables, suffers from this non-locality. This holds, because
neither the current theory nor the alternative ones are responsible
for non-locality. We will try to prove that the idea of two,
non-interacting, distant particles in zero spin state "together" is
solely responsible for the whole novelty of non-locality.

Let single particles in zero spin state decay into two correlated
fermions, the fermions traveling northwards and the fermions
traveling southwards, or let a positronium with zero angular
momentum decay into its constituents, the electron and the positron.
When performing a spin measurement on the "north" $N$ fermions along
$z-$axis, we find $N/2$ in $|+z\rangle$ spin state. We ignore the
others. Performing a new spin measurement along $x-$axis, we find
$N/4$ in $|+x\rangle$ spin state. We ignore the others. Performing
again a spin measurement along $z-$axis, we find $N/8$ fermions in
$|-z\rangle$ spin state. We ignore the others. Let us now see the
state of the "south" fermions correlated with the above $N/8$ north
fermions. The north fermions have passed from $|+z\rangle$ to
$|-z\rangle$ spin state. So, the correlated south fermions have
passed from $|-z\rangle$ to $|+z\rangle$ spin state, so energy and
angular momentum were exchanged between the north set of $N/8$
fermions and the south one.

Large numbers of particles, statistical behavior and "strong" or
"weak" versions of entanglement do not affect our discussion. Let us
imagine a single particle decaying into two correlated fermions, the
north and the south one. There is 12.5 per cent probability the
north fermion to be caught in $|+z\rangle$ spin state at the first
measurement, and in $|+x\rangle$ spin state at the second
measurement, and in $|-z\rangle$ spin state at the third
measurement. So, at least for this pair of fermions, energy and
angular momentum transferred from the one fermion to the other,
because the south fermion was in $|-z\rangle$ initially and in
$|+z\rangle$ finally, though no measuring device interacted with it.
(In fact, one can easily prove, by taking into account all the
possible outcomes of the experiment, that this angular momentum
transfer takes place with 50 per cent probability.) This is the very
meaning of non-locality. How did this energy penetrate walls to pass
from the one fermion to the other? How did it travel the long
distances of the experiment? What kind of Hamiltonian is this, which
does not decrease with distance? What kind of Hamiltonian is this,
which would prevent Newton from his first law proposition ("a remote
body can not be affected strongly by other bodies, so the total
force on it is negligible and consequently it can be taken as an
inertial frame of reference")? Is there any field (equivalently
force) that makes the correlated particle to reverse its spin, or
there is not such a thing? If the second holds, the angular momentum
(and/or energy and momentum in other cases) changes without a
"reason", i.e.: without a field(/force) enforcing this change, so we
should wave Lagrangian/Hamiltonian and Newtonian formulations
goodbye. Newton lied when he wrote $dL/dt=\tau$, here we do have
$dL/dt$ (and the uncertainty relations have nothing to do with our
discussion), but we have no $\tau$, namely there is neither a
measuring device in the vicinity of the overturned particle nor a
field to cause this change in the particle spin. Schr\"{o}dinger, on
the other hand, lied when he postulated that the evolution of some
system is determined by his equation, this is not true not only when
one measures this system, but also when one measures some other
systems somewhere in the world. In other words, when Schr\"{o}dinger
claimed that the correlated particle, the one for which no
measurement is performed, is in $|-z\rangle$ spin state, and,
because $\hat{H}=0$, it remains in this state, just lied. A magic
measurement in another particle inevitably changes $|-z\rangle$ into
$|+x\rangle$ or $|-x\rangle$. After all, nobody knows when
Schr\"{o}dinger equation holds, because nobody knows when someone
else decides to perform its measurements, destroying the validity of
the equation for the correlated particles. If there exists a field
"joining" together the entangled particles, carrying the above
amounts of energy and angular momentum, what sort of physical object
is this, that does not decrease with distance? Is there an
infinite-energy source which covers the total space? Or the field
has not to cover the total spacetime, but it is so "smart", so as to
recognize which two particles are really entangled and orientate
itself from the one particle to its entangled partner, like a smart
and unbelievably reliable messenger? If I had a positron at the
galaxy of Andromeda and its correlated electronic partner here in
Earth, mixed with many other electrons within a metal, how smart
should this energy be, coming from the positron, so as to recognize
the entangled electron within the metal, and overturn this electron
and not another one? Or the energy has not to find out the
correlated particle but overturns the first electron that happens to
meet when traveling? Do my electrons suffer from such unwarned
spin-flips, because my neighbour uses to produce correlated fermions
and to overturn the set of the positrons? Such a hypothesis may be
easily tested experimentally, but has nothing to do with genuine
non-locality. So, the energy coming from the one, the measured,
fermion is clever enough to distinguish between an "entangled"
fermion, and $10^{23}$ "un-entangled", identical fermions. Does the
entanglement mark the correlated particles with some number/color or
is it something like a disease? In a somewhat different case, I use
$10^{23}$ hiding places (equal in number stars from our galaxy or
others) to hide $10^{23}$ positrons. The electrons correlated to the
above positrons consist the fermion sea of my wristwatch. As it is
well known, these positrons are not identical because their wave
functions do not overlap, \cite{beiser,eisberg}. So, the electrons
in my watch are not identical (though their wave functions overlap,
and they overlap because Schr\"{o}dinger equation seems to hold, at
least under "normal" conditions, with measurements neither here, nor
anywhere in the world for fear of happening upon a correlated
partner), they have a label, the name of the star that I hide the
correlated positron. So, quantum statistics is gone. No "weak"
version of correlation can be applied: The positrons do not mix, I
may label them one by one and, consequently, I may label the
electrons one by one, so, the energy/angular momentum coming from
the a-Taurus positron can not be absorbed by the b-Scorpio electron,
this energy will be absorbed only by the a-Taurus electron (i.e.: by
the electron correlated to the a-Taurus positron).

The situation described in the above paragraph is not a mere
scandal; It is an obvious dilemma: Either non-locality, or Physics.

An energy-momentum non-conservation has been proposed by some
authors, not arising from the uncertainty relations, because the
change in angular momentum of the second particle is permanent,
after the third measurement in the first particle. This
non-conservation is impossible to arise from the interaction between
the particle and the measuring device for two reasons: No
measurement is performed in the second particle, and no energy,
momentum or angular momentum can be exchanged between the devices
and the system of the particles, as it is supposed to be in zero
spin state (unless it is \emph{not} in zero spin state, as we claim
here), and this state never exchanges angular momentum with anything
in the world.

Some authors believe that a measurement on one of the correlated
particles affects instantly its partner, changing for example its
spin. Others try to formulate a covariant theory for these
measurements, claiming that the energy or other measurable
quantities should travel with the velocity of light. Few believe in
retrospective signals. These arguments do not affect our discussion.
The above peculiarity in energy and angular momentum transfer, what
we called non-locality, remains, whichever is the frame of our work,
Newtonian or Einsteinian. And, after all, our problem is this very
telepathy (much "stronger" than the rumored human one, because in
physics there are physical quantities that travel and not simply
some "information" between different persons), not its covariance or
its retrospective character.

Some believe that non-locality comes from the hidden variable
quantum mechanics. On the other hand, it is a common belief,
especially among quantum information theorists, that the orthodox
theory is fundamentally non-local. We said nothing here about hidden
variables, stochastic formulations of quantum theory, quantum
potentials or pilot waves. We said nothing about which model, the
orthodox, or an alternative one, we prefer. We said nothing about
the other, the "epistemological", problem, if the different quantum
mechanical models are fundamentally different theories, or their
existence just consists a secondary matter of interpretation. We
just reproduced the results of the orthodox theory, also reproduced
by the experiments and by hidden variable theories (at least by
serious competitors of the orthodox theory) and found that
non-locality, in the sense that energy and angular momentum is
transferred, is present.

\section{How to reintroduce locality}

Our purpose is not to measure the violation of Bell inequalities in
ensembles of entangled particles or in isolated pairs of entangled
particles. Our purpose it to pose the question of the possibility of
the zero spin state for remote particles (never phrased, up to our
knowledge) and to propose some ideas to test this possibility
experimentally. The zero spin state assumption for remote particles
is a necessary one for the proof of Bell inequalities and leads,
according to the discussion of the above paragraph, to a series of
peculiar phenomena. It is clear that the usual theorem
"realism+determinism"$\Rightarrow$"Bell
inequalities"$\Rightarrow$"non-locality", which can be found in
several versions in any review of the topic, is deceptive. A more
formal formulation of the theorem, according to our present ideas,
is: "realism+something else that one likes (usually
determinism)+\emph{zero spin state assumption for distant,
non-interacting particles}"$\Rightarrow$"non-locality". I hope that
the quite tiring discussion of the previous section made as plain as
a day that the right formulation is "\emph{zero spin state
assumption for distant, non-interacting
particles}"$\Rightarrow$"non-locality", because no realism or
determinism or anything like that, but only the above assumption,
slipped into our discussion. Up to our knowledge, nobody rejects the
existence of zero spin state for distant, non-interacting particles.
They all regard it as a simple truth.

How can two particles be in zero spin state? In the positronium
example there is a Hamiltonian interaction between them. The same
holds true for the zero spin mesons and their constituent quarks.
They are not automatically in zero spin state. So, when the
constituents of the positronium are at distance, they can not be in
zero spin state, because such a Hamiltonian does not exist.

In the present article we claim that two particles coming from the
decay of a single zero spin object are not in zero spin state, but
in $|+n\rangle_1|-n\rangle_2$ spin state, where $n$ is a random
vector in the usual space. The angular momentum is conserved when
the initial particle decays, but it is not conserved when
measurements are performed either to one or to both of the product
particles. To experimentally test this assumption we should use
single pairs of correlated particles. As one can easily prove, for
ensembles of correlated particles, both $|+n\rangle_1|-n\rangle_2$
and $|0\rangle$ spin state show similar behavior and, thus, can not
be distinguished experimentally.

We remind that this energy/angular momentum non-conservation before
and after a measurement is widely known in quantum mechanics, as
measuring devices and particles exchange energy (expect for some
rare states, like $|0\rangle$ one). The above particles are in an
eigen-state of the $L_n$ operator and they will remain in this state
if we perform a spin measurement along $n-$axis. But if we perform a
spin measurement along an $n'-$axis, the angular momentum will not
be conserved for the measured system.

So, when measuring the angular momentum of the
$|+n\rangle_1=a|+z\rangle+\sqrt{1-a^2}|-z\rangle$, $0\leq a\leq1$,
fermion along $z-$axis, there is $a^2$ probability to find it in
$|+z\rangle$ spin state and $(1-a^2)$ to find it in $|-z\rangle$
spin state. When measuring the angular momentum of the
$|-n\rangle_2=\sqrt{1-a^2}|+z\rangle+a|-z\rangle$ fermion along
$z-$axis, we find $(1-a^2)$ probability for the $|+z\rangle$ spin
state and $a^2$ probability for the $|-z\rangle$ spin state,
independently of the result of the measurement in the other fermion.
Why "independently"? Because $|+n\rangle_1|-n\rangle_2$ is not an
eigen-state of the $L_z$ operator, as the zero spin state is. So
there is no energy or angular momentum transfer between the two
product particles, thus there is no correlation, and thus there is
no need to introduce non-locality. The only energy and angular
momentum transfer takes place between the particles and the
measuring devices.

The above paragraph shows a straightforward way to test our claim
for the $|+n\rangle_1|-n\rangle_2$ state of the particles that arise
from the decay of a zero spin particle. If the two fermions continue
to be in zero spin state, i.e.: if they are "entangled", after their
release from the initial particle, any measurement of their angular
momentum along $z-$axis, will give two possible results with equal
probabilities: either $|+z\rangle_1|-z\rangle_2$, or
$|-z\rangle_1|+z\rangle_2$. The possibility for both particles to be
in $|+z\rangle$ or in $|-z\rangle$ spin state is zero. But if the
two particles are not correlated after their release, and, as we
claim here, they are in $|+n\rangle_1|-n\rangle_2$ spin state, then
there is $a^2(1-a^2)\neq0$ probability for
$|+z\rangle_1|+z\rangle_2$ spin state and $a^2(1-a^2)$ probability
for $|-z\rangle_1|-z\rangle_2$ spin state. (The only case that the
quantity $a^2(1-a^2)$ equals to zero would be our misfortune to
accidentally choose the axis $z$ of measurement the same with
$n-$axis. So, we may use at least two different axes for the same
pair of particles.)

\section{Some experimental hints to test the zero spin state assumption for remote particles}

\begin{figure}
\centering
\includegraphics{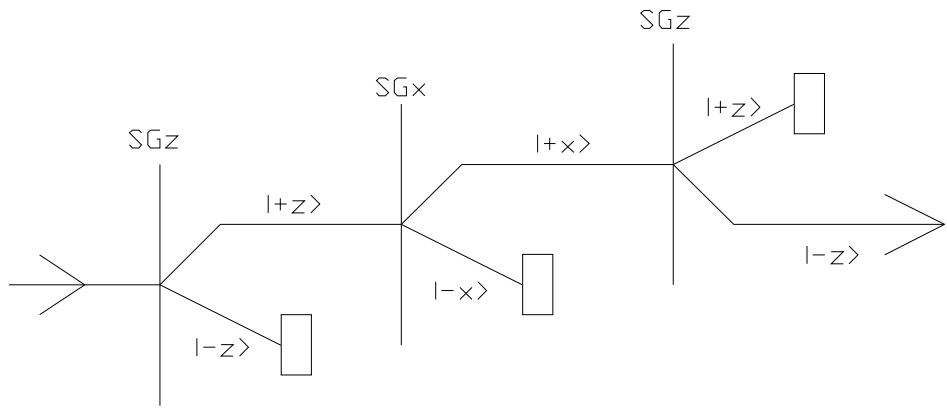}
\caption{A possible path for a muon through three Stern-Gerlach (SG)
devices. If the muon follows the obstacle-free trajectory (1/8
probability), then its spin orientation changes from $|+z\rangle$ to
$|-z\rangle$.} \label{figure1}
\end{figure}

Let us now see the experiment, in figure \ref{figure1}. A pion can
decay to a muon and a muonic neutrino: $\pi^+\rightarrow
\mu^++\nu_{\mu}$. Let this muon travel through the first Stern
Gerlach device (SGz) orientated along $z-$axis. There is 1/2
probability for the muon to be in $|+z\rangle$ spin state. The
second SG device is orientated along $x-$ axis. Let this measurement
give $|+x\rangle$ for the muon spin. The muon and its neutrino
partner will be in zero spin state, equivalently, they will be
correlated, if and only if the results of the third measurement, in
the last SG device, are all $|+z\rangle$. If there are any muons
escaping from the whole system of devices, following the trajectory
of figure \ref{figure1}, namely, if there are any muons being caught
in $|-z\rangle$ spin state at the end of the third measurement, then
the muonic neutrino, being initially in $|-z\rangle$ spin state, is
enforced to turn into $|+z\rangle$ spin state at the end of the
experiment, to maintain the zero total angular momentum of the two
entangled particles. So: Either the neutrino changes its helicity
(impossible, unless it is not massless), or the neutrino makes a
U-turn, (reversing both momentum and angular momentum, so as to
maintain its helicity) and the momentum is not conserved, as some
authors claim (this energy-momentum non-conservation is not over a
space or time interval imposed by the uncertainty relations, its an
absolute, permanent non-conservation), or the neutrino is
transformed into an antineutrino and the momentum is conserved but
not the muonic lepton number, which is, after all, more possible
than the energy-momentum non-conservation (we remind the above
dilemma, non-locality or Physics), or something much simpler
happens: The two remote particles are not correlated/entangled, and
the neutrino is not obligated to change its spin from $|-z\rangle$
to $|+z\rangle$ state, because an inverse change happened, somewhere
in the universe, to a muon, supposed to have exchanged vows of
eternal entanglement with its neutrino better half.

\begin{figure}
\centering
\includegraphics{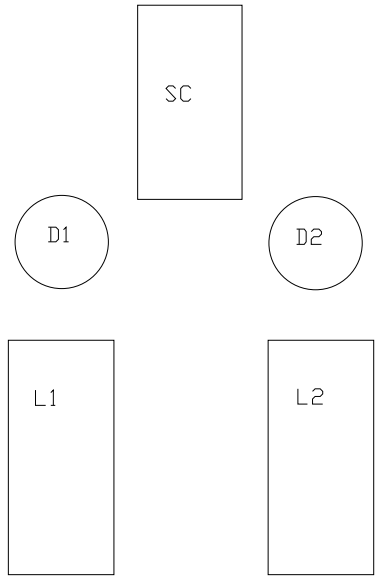}
\caption{The superconductor (SC) can provide Cooper pairs in singlet
spin state, tunneling to the two quantum dots (D1, D2). Each of the
two electrons may follow by tunneling one of the leads, L1 or L2.
According to our present ideas, the two particles continue to be in
$|0\rangle$ spin state, even when traveling along different leads.
Here, we proposed that the two particles are in
$|+n\rangle_1|-n\rangle_2$ spin state, after tunneling in the two
leads, they are in $|0\rangle$ spin state only within the
superconductor. We need some Stern-Gerlach devices to test the real
spin state of the two particles after tunneling.} \label{figure2}
\end{figure}

We remind that for a free (i.e.: un-entangled) muon or other
fermion, the probability, according to both orthodox quantum theory
and to any serious hidden variable alternative, to follow the
trajectory of figure \ref{figure1} is 1/8. So, if no muon follows
the trajectory of figure \ref{figure1}, then entanglement exists.
But if N/8 muons follow this trajectory, either something strange
and unnatural, like that described above, happens with poor
neutrino, or entanglement does not exist.

One may also use Cooper pairs in superconductors and generally the
technology of electron spin entanglers in condensed matter physics,
\cite{cond1,cond2,cond3,cond4,cond5,cond6,cond7,cond8,cond9,cond10,cond11}
to test the existence of entanglement for a single pair of
particles. The experiment we propose in figure \ref{figure2} needs a
superconductor, two quantum dots and two usual leads, as in, for
example, \cite{exper1,exper2,exper3,exper4}. The device is quite
simple. Two electrons, forming a Cooper pair within the
superconductor, can tunnel, by means of Andreev tunneling, to two
quantum dots, each electron to different quantum dot. Then, the
electrons may tunnel from the quantum dots to two normal leads, and
follow, each one, two distinct trajectories. A system of
Stern-Gerlach devices can pick each one of the two electrons and
test if they are in singlet (as the Physics community regards) spin
state (namely, if they are entangled), or in
$|+n\rangle_1|-n\rangle_2$ state, with no sort of correlations
between them.

In conclusion, we showed that entanglement, leading to a peculiar
transfer of measurable quantities from the one entangled particle to
the other, is the result of the assumption that some remote
particles are in $|0\rangle$ spin state "together". We proposed some
simple experimental procedures to test this assumption.

\end{document}